%Paper: astro-ph/9506116
%From: Basilio Santiago <santiago@ast.cam.ac.uk>
%Date: Thu, 22 Jun 1995 16:33:43 +0100 (BST)

\magnification=\magstep1
\hsize=15truecm
\vsize=22truecm
\hoffset=0.5truecm
\def\double{\baselineskip=0.9truecm}

\raggedbottom
\font\text=cmr10
\text
%                    ASTROPHYSICAL JOURNAL FORMAT

\def\apj    {{\it Ap.~J.~}}
\def\apjl   {{\it Ap.~J.~(Letters)~}}
\def\apjs   {{\it Ap.~J.~Suppl.~}}

\def\aj     {{\it A.~J.~}}
\def\aa     {{\it Astr.~Ap.~}}
\def\mnras  {{\it M.N.R.A.S.~}}

\def\araa   {{\it A.R.A.A.~}}

\def\nature {{\it Nature~}}

\def\ref {\noindent}
\def\v#1  {{\bf {#1}~}}

\def\etal {{\it et al.~}}

\double

\vglue 2cm

\centerline {\bf The Globular Clusters in M87: A Bimodal Colour
Distribution $^1$}

\vskip 3cm

\centerline {Rebecca A. W. Elson and Basilio X. Santiago }
\centerline {\it Institute of Astronomy, Madingley Road, Cambridge CB3 0HA,
 England }
\centerline {elson or santiago @ast.cam.ac.uk}
\vskip 4cm
\noindent
Submitted to {\mnras} May 10 1995.

\noindent
Revised June 16 1995.

\vskip 3cm
\noindent
$^1$Based on observations obtained as part of the Medium Deep Survey.

\vfill\eject

\centerline {\bf Abstract}

We present $V$ and $I$ photometry for $\sim 150$ globular clusters with
$20 < V < 27$
$\sim 2.5$ arcmin from the centre of M87, the cD galaxy in the
Virgo cluster.  The data were acquired with the Hubble Space Telescope which,
with effective resolution 0.1 arcsec, allows us for the first time to
distinguish between globular clusters and most background galaxies on the
basis of morphology, and to obtain accurate photometry even at faint
magnitudes.  The $(V-I)$ distribution of the clusters is
clearly bimodal, implying a corresponding bimodality in the metallicity
distribution, with peaks at [Fe/H]$\approx -1.5$ and $-0.5$.
We also find  that the brightest clusters are predominantly metal poor,
while the fainter clusters are divided roughly eaually between metal
poor and metal rich.
Our sample is essentially complete
to $ V\approx 25$,  a full magnitude past the expected peak of a ``universal"
luminosity function, and the observed luminosity function
is well represented by the standard  gaussian distribution
with $<M_V>=-7.2$ and
$\sigma = 1.4$.


\vskip 2cm
\noindent {\bf 1. Introduction}
\smallskip

Detailed studies of globular cluster systems in various types
of galaxies are beginning to yield important clues regarding the formation
of both the globular clusters and their host/parent galaxies.  (For
recent reviews see Harris 1991; 1995).
In particular, the large populations of clusters often associated with
elliptical galaxies have prompted the suggestion that clusters
form during mergers of the subunits which may have gone into building the
galaxies (cf. Ashman and Zepf 1992).  This possibility has gained popularity
recently in light of the
discovery
of objects which appear to be young globular clusters
in NGC 1275, NGC 3597, NGC 7252, and He 2-10, all
disturbed galaxies which have probably undergone recent
mergers (Holtzman \etal 1992;
Lutz 1991;
Whitmore \etal 1993; Conti and Vacca 1994).
However, since field stars would also form during mergers,
such a scenario may have trouble accounting for the higher specific
frequencies of clusters in many elliptical galaxies (cf. Harris 1995).

One signature of
a merger origin for globular clusters, or indeed of any scenario in which
clusters form in successive bursts,
might be a bimodal or multimodal
metallicity distribution (Zepf and Ashman 1993), which
should manifest itself as a bimodal or multimodal colour distribution
in indicies such as $(B-I)$ or $(V-I)$.
Zepf and Ashman show
histograms of metallicity for globular clusters in
two galaxies, NGC 4472 and
NGC 5128, both of which are bimodal. NGC 5128 is a peculiar elliptical
galaxy with a dust lane and a jet, and NGC 4472 is an E/S0 shell
galaxy with CO emission.   On the other hand,
Ajhar, Blakeslee
and Tonry (1994) present $(V-I)$ histograms for globular clusters associated
with 10 elliptical and S0 galaxies in the Virgo and Leo clusters and
find that, while  the histograms
differ substantially from galaxy to galaxy, none shows
evidence for bimodality.

M87 (NGC 4486), the central giant elliptical in the Virgo cluster,
was not among the galaxies studied by Ajhar \etal because its
high
specific number of clusters made it unrepresentative
of elliptical galaxies in general.  Its jet also makes it atypical.
However,
understanding the origin of its prodigious population of clusters,
whether primordial, by accretion of companion galaxies with their attendant
clusters, or in more recent episodes of cluster formation in mergers or
cooling flows, is an important quest.

Recently, Lee and Geisler (1993) have used Washington system photometry
to estimate metallicities for about 400 M87 clusters at radii $\sim 1-4$
arcmin.  They find a broad distribution of values with a mean of
[Fe/H]$\approx -0.9$.
Couture, Harris and Allwright (1990) present a $(V-I)$ histogram for a
sample of 106 M87  globular cluster candidates
brighter than $V\sim 23.4$.  They also find a
broad distribution of colours, with $<V-I> \approx 1.0$, which they estimate
corresponds to [Fe/H]$\approx -1.0$. Neither of these studies reveals
evidence for bimodality in the metallicity distribution.

In this paper we present $V$ and $I$ photometry for a sample of $\sim150$
globular clusters with $V < 27$, observed with the Hubble Space
Telescope (HST).  HST enables us to obtain accurate colours even at
faint magnitudes, and
is unique in allowing us to distinguish between globular clusters and most
background galaxies on the basis of morphology, and thus to be assured of a
relatively uncontaminated sample.
Our observations and sample selection are described in Section 2, and our
results are presented in Section 3.

\vskip 1cm
\noindent {\bf 2. Observations }
\smallskip

Our study is based on images of a field located $\sim 2.5$ arcmin from the
centre of M87,
obtained with HST's refurbished Wide Field Camera (WFC-2) on 1994 August 4,
as part of the Medium Deep Survey Key Project (cf. Griffiths
\etal 1994).   The images were acquired in parallel mode, in fine lock.
The F814W ($\sim$ Cousins $I$)
and F606W ($\sim $ Cousins $V$) filters were used.  Two
1000 second exposures were obtained with each.
The gain is 7.0 $e^-$/ADU, and the readnoise is 7 $e^-$.
The image scale for the WFC-2
is 0.10 arcsec/pixel.
Adopting a distance modulus for M87 of 31.0 based on recent observations
of Cepheids (Pierce \etal 1994, Freedman \etal
1994) gives a pixel size of 8.7 pc.
A typical globular cluster would have core radius less than 10 pc
and tidal radius of order 50 pc;  in  our relatively shallow exposures
only the cores are visible, and the clusters are therefore unresolved.
Eliminating the unexposed edges of each of the three 800x800 pixel WFC-2 chips
gives a field of view with total area 4.7  arcmin$^2$.
(To maintain homogeneous selection criteria, the Planetary Camera image,
with smaller scale and area only 0.3  arcmin$^2$ was not used.)

The individual images in each colour were
coadded using an IRAF package developed by K. Glazebrook.
In order to remove cosmic ray events,
the smallest of the two pixel values at each position in each chip
was adopted. For our 1000 second exposures, only about 2\%
of cosmic ray events should escape elimination by this process,
affecting only 0.03\% of the pixels (WFPC-2 Instrument Handbook version 2.0).
Since the individual frames are not dithered,
hot pixels could not be removed by coadding.  Instead,
they were identified and flagged using software
kindly provided by N. Tanvir.
Figure 1 shows a 255x255 pixel subsection of the coadded F606W Chip 3 image.

\medskip
\noindent
{\it 2.i. Object detection and photometry}

Objects were detected automatically using the Cambridge APM software (Irwin
1985).
The adopted detection threshold was 1.5 times
the standard deviation of the background, which
corresponds to a surface brightness in $I_{814}$ of
23~mag~arcsec$^{-2}$. In order to
eliminate cosmic ray residuals, chip defects, noise spikes,
and very faint unresolved
galaxies, a minimum diameter of 0.3 arcsec was also imposed.

Magnitudes were measured using aperture photometry with an
aperture of radius 2 pixels.
The full radius of the
point spread function
is $\sim 25$ pixels (Holtzman \etal 1995), so an aperture correction was
required to derive total magnitudes.
This was determined to be 0.38 mag from observations
of isolated compact objects.
Clearly any resolved galaxies would require larger
aperture corrections, depending on their surface brightness profiles, and their
magnitudes may be underestimated here.  This does not pose a problem for us
since our only interest in the galaxies is in
eliminating them from our sample.
The adopted sky brightness for each object
was the mode in an annulus at radii 30-35 pixels.
Photometric
zero-points are 21.67 for the F814W filter, and 22.84 for the F606W filter
(Holtzman \etal 1995).

The instrumental HST magnitudes were converted to Johnson-Cousins magnitudes
($V,I$) using the
relations

$$ I = I_{814} + 0.001 - 0.126 (V_{606}-I_{814}) + 0.047 (V_{606}-I_{814})^2 $$

\noindent
and

$$ V = V_{606} - 0.033 + 0.252 (V_{606}-I_{814}) + 0.004 (V_{606}-I_{814})^2 $$

\noindent
These include conversions from space-based to ground based WFC-2 magnitudes
(Holtzman \etal 1995) and from ground-based WFC-2 magnitudes to the
Johnson-Cousins system (Harris \etal 1991).  The $V$ conversion involves
a $(B-V)$ colour term for which we adopted the relation
$(V-I)=1.29(B-V)+0.12$ from Couture \etal (1990).  This is only valid
for M87 globular clusters, and there is significant scatter
about this relation, which will introduce uncertainties in the conversion.
Further uncertainties are introduced because the Harris \etal transformations
are themselves based on observations of individual dwarfs, and may not
therefore be valid for globular clusters, where the integrated light tends
to be dominated by giants.
The transformations are therefore tentative, and
we present our results in both the HST instrumental and
Johnson-Cousin systems.
The difference between the HST and Johnson-Cousins $I$ magnitudes is less than
0.08 mag.  In $V$, the difference ranges from 0.05 mag
at $(V_{606}-I_{814})=0.4$
to 0.25 mag at  $(V_{606}-I_{814})=1.1$.
A colour-magnitude diagram for 238 objects with $V_{606}<27$ is shown in
Fig. 2.  Errors range from $\pm 0.02$ mag at  $V_{606}=22.5$  to $\pm 0.05$
at $V_{606}=23.5$.

\medskip
\noindent
{\it 2.ii. Elimination of stars and galaxies}

We now wish to eliminate stars and
galaxies from our sample.  We begin by restricting our sample to objects
with colours consistent with those of globular clusters.  Referring
to Fig. 5 of Couture \etal (1990),
reasonable limits are $0.6 < (V-I) < 1.4$, which corresponds to
$0.47 < (V_{606}-I_{814}) < 1.08$.  Adding $\pm 0.1$ to account for
scatter in the observed colours, we make our initial selection at
$0.37 < (V_{606}-I_{814}) < 1.18$.  These limits are
indicated with dashed lines in Figure 2.
This gives a sample of 187 globular cluster candidates.

We can estimate the expected number of foreground stars and background
galaxies using the Medium Deep
Survey database.   A study of Galactic stars in 13 high-latitude WFC-2 fields
(Santiago \etal 1995)
indicates that we should expect about one star per magnitude bin
per field, so at $20 < V < 26$, about six stars altogether, or two per chip.
Since the cores of the globular clusters are unresolved, it is not
possible to distinguish them from stars on the basis of the appearance
of the images.  However the fraction of stars expected is only
$\sim 4\%$ of the final sample, and
this level of contamination will not affect any of our results
significantly.

Unlike stars, most background galaxies can be eliminated by
visual inspection of
the images.  Edge-on or elongated galaxies and lumpy or low
surface brightness
objects are easily distinguished from globular clusters as can be seen in
Figure 1.
We identified 40 galaxies among the 187
globular cluster candidates.  This is consistent with the number expected
based on the same 13 high latitude
Medium Deep Survey fields discussed above.
Eliminating these galaxies from our sample leaves a total of 147 objects.

Distant early-type galaxies are not morphologically distinguishable from
globular clusters and may still contaminate our sample. However,
ellipticals and compact galaxies account for only about 20\% of the
Medium Deep Survey galaxy
population brighter than $I =22.0$ ($V\approx 24$)(Griffiths \etal 1994)
and even less than that at fainter magnitudes and with $0.6 < (V-I) < 1.4$
(Driver \etal 1995, Glazebrook \etal 1995).
Thus, these objects should constitute less than 8\% of the final sample,
giving a total
contamination by stars and galaxies of less than $12\%$.

\vskip 1cm
\noindent {\bf 3. Results }
\smallskip
\noindent
{\it 3.i. Colour distributions}

Figure 3a shows a histogram of ($V_{606}-I_{814}$) values for the 187
globular cluster candidates selected by colour, and for the final sample
of 147 globular clusters.  The objects identified as galaxies tend
to lie in the wings of the distribution.
Both histograms are clearly bimodal, with peaks at ($V_{606}-I_{814})\approx
0.6$ and 0.8.

Figure 3b shows a histogram for the final sample of 147 globular
clusters, with magnitudes transformed to the Johnson-Cousins system.
Superposed is the  histogram from Fig. 5 of Couture \etal for 106
M87 clusters with $V<23.4$.  There appears to be a systematic offset
in the sense that our histogram is shifted bluewards by $\approx 0.15$
mag.  This is probably due to uncertainties in our transformation from
the HST to Johnson-Cousins magnitudes, discussed above.  Given
that the transformation is based on observations of dwarfs, not
globular clusters, and that there is a $(B-V)$ term which must be
approximated by a relation which is known to have significant scatter,
a systematic uncertainty of 0.15 mag in $(V-I)$ would not be unlikely.
The important feature of our histogram is its bimodality,
and this is not affected by the transformation.

As discussed above, such bimodality has been observed in
the globular clusters systems of several
other galaxies. However, it has not been noted in previous studies
of the M87 cluster system.
As Couture \etal point out, studies using $(B-V)$ or $(g-r)$ indicies
would not be expected to show such bimodality because the span
of wavelengths is too narrow. However, their $(V-I)$ histogram might be
expected to show bimodality.  To compare our results directly with theirs,
Fig. 3c shows a histogram for the 76 brightest  clusters in our
sample, with $V<23.5$, shifted redwards by 0.15 mag.  Again, the
histogram from Couture \etal is superposed.
The latter does
show some suggestion of bimodality with peaks similar to ours,
although it is also consistent with a
single broad distribution.
The errors in Couture \etal's photometry
are $\pm 0.2$ mag at $V=21.5$ and $\pm 0.3$ mag
at $V=22.5$.
In comparison, our errors are an order of magnitude less:  $\pm0.02$ mag
at $V=22.5$ and $\pm 0.05$ mag at $V=23.5$.
It is likely that their errors are too large
to fully resolve the bimodality evident in our sample.
Lee and Geisler (1993) present a histogram of values of [Fe/H] for
M87 globular clusters, derived from Washington system narrow band photometry.
They too find a broad distribution of values ranging from [Fe/H]$\approx -2$
to $0.5$. They estimate their errors in the photometric index to
be $\pm 0.06$, which translates into an error in [Fe/H] of
$\pm 0.15$ dex.  Again, these uncertainties
may be too large to resolve a  bimodality.

A difference in colour among clusters could in principle be due
to a difference in age, in the sense that younger clusters would have
systematically bluer colours.  However as Couture \etal note, $(V-I)$
colours are relatively insensitive to age: an age difference of, for
example, 2 Gyr would not affect $(V-I)$
significantly.  The difference is far more likely to be due to a difference
in metallicity.
Using the relation given by Couture \etal between colour and metallicity,
we estimate that
the blue peak in Fig. 3c corresponds to [Fe/H]$\approx -1.5$,
and the red peak to
[Fe/H]$\approx -0.5$.
These values are
consistent with the peaks at $-1.7$ and $-0.5$ found
by Zepf and Ashman (1993) in NGC 4472, and are slightly more metal poor
than those in NGC 5128, at $-1.2$ and $-0.1$.
Our results therefore suggest that there were at least two distinct
episodes of cluster formation in M87, or that clusters were
accreted from parent galaxies with different metallicities.

Interestingly, there appear to be relatively more metal poor (bluer)
clusters in our brighter sample compared to our fainter sample.
Figure 3d shows histograms
for the clusters with $V<23.5$ ($N=76$),
and with $23.5 < V < 25$ ($N=64)$.
Among the bright clusters, the blue peak is dominant, whereas among
the fainter clusters, the red peak is dominant.  Specifically,
for the bright sample, $72\%$ of the clusters have $(V-I)<0.9$ while
$28\%$ have $(V-I)>0.9$.  For the fainter clusters, the fractions
are $42\%$ and $58\%$ respectively.

This  trend is also present in the sample prior to converting the
instrumental magintudes to the Johnson-Cousins system,
and is therefore not the result of any systematic uncertainties  in the
conversion. Nor is it a result of any colour-dependent
incompleteness. As Fig. 4 (below) illustrates,
our sample should be complete to $V=25$.  If, however, we adopt a more
conservative completeness limit at $V=24$ and repeat the above analysis,
dividing the sample at $V=22.8$, we still find the same trend.
Finally, we ask whether the trend could be the result of contamination
of our sample by faint, red, unresolved galaxies.  MDS results for a
sample of ``compact" and elliptical galaxies with $V<23$ show no systematic
shift redwards
at fainter magnitudes in the range of colours considered
here.  Morphological classifications at $V>23$ were not possible,
however we estimate above that contamination of our sample by compact
galaxies indistinguishable from globular clusters is about $8\%$, or
about 12 galaxies.  If we adopt the extreme assumption that all 12
possible interlopers
are faint and red and remove them from the sample accordingly,
then the fraction of blue and red clusters in the fainter sample
changes from $42\%$ and $58\%$,
to $52\%$ and $48\%$.  The ratios for the bright clusters remain unchanged.
Thus, even in this case, the trend of brighter clusters being predominantly
blue or metal poor compared to the fainter clusters  persists.

Such a trend has never before been noted in any galaxy. The likely reason for
this is that to resolve it requires very accurate
photometry at faint  magnitudes.
The peaks in our colour distribution are
separated by only 0.2 mag, and the median colours of the bright
and faint samples differ only by 0.1 mag.

If real, the trend illustrated in Fig. 3d
suggests that lower metallicity environments favour the
formation of more massive clusters, while higher metallicity environments
tend to produce smaller clusters.
Harris and Pudritz (1994) present a model for cluster formation in
which clusters form in the cores of supergiant molecular clouds,
with mass proportional to the size of the progenitor cloud.  If at a later,
more chemically enriched epoch, the progenitor clouds had been broken
into smaller pieces by tidal forces or through cloud-cloud collisions,
then one would expect  the more metal rich clusters to be smaller.  Indeed,
a somewhat analagous trend is seen in the Large Magellanic Cloud,
where none of the (more metal rich)
younger clusters are as massive
as the oldest (more metal poor) clusters.  (Even if the younger clusters
are currently as massive as the oldest clusters,
they will lose a significant
fraction of their mass through stellar evolution and evaporation. Thus,
their predicted final masses are smaller than the masses currently observed
among the oldest clusters.)

\medskip
\noindent
{\it 3.ii. Luminosity functions}

Figure 4 shows a luminosity function for the 147 globular clusters in
our final sample.
Superposed is a gaussian distribution with $<V>=23.8$ ($<M_V>=-7.2$),
and $\sigma=1.4$, the values given by Flemming \etal (1995) for
the M87 globular
cluster system, normalized to the total number of clusters  with
$V<24$ ($N=100$).  At this is the magnitude we expect our sample to be
complete, based on results from Santiago \etal (1995) who make use
of simulations of HST images
with various signal-to-noise ratios to determine completeness as a function
of magnitude in images similar to these.
The gaussian function matches the data very well for $V < 25$.
Fainter than this our sample rapidly becomes incomplete.

Integrating the gaussian distribution implies a surface density of 38
clusters per square arcmin.  Harris (1986) derives a surface density
profile for the M87 globular cluster system which implies
20 clusters per square arcmin at a radius of 2.5 kpc.  His estimate
of the limiting magnitude of his sample is uncertain, but corresponds
roughly to $V=23.5$.  To this limit we observe 16 clusters per square arcmin,
in good agreement with Harris' value.
Thus, we confirm his result that the globular cluster system of M87 is
less centrally concentrated than the underlying galaxy.

\vskip 1cm
\noindent
{\bf 4. Conclusions }

We have presented accurate $V$ and $I$ photometry for
$\sim 150$ globular clusters
in M87.  We find a colour distribution which is clearly bimodal, with
peaks corresponding to metallicities of [Fe/H]$\approx -1.5$ and $-0.5$.
Among the brighter clusters,  with $20 < V < 23.5$, metal poor clusters
constitute $72\%$ of the sample.  Among the fainter clusters, with
$23.5 < V < 25$, metal poor clusters constitute only $42\%$ of the sample.
These results suggest that there were at least two distinct episodes
of cluster
formation in M87, or that the clusters were accreted from parent galaxies with
different mean metallicities.  They suggest further
that metal poor environments may favour the formation of
more massive clusters, while less massive clusters form preferentially
in more metal rich environments.
Finally, our sample is complete to $V\approx25$,
a full magnitude past the expected peak of a ``universal" luminosity function,
and we find that our observed luminosity function is well represented
with the standard gaussian function, with $\sigma = 1.4$.

\vskip 1cm
\noindent
{\bf Acknowledgements}

We thank Bill Harris for many useful comments and suggestions.  Tables of
photometry are available from the authors upon request.
\vfill\eject
\noindent{\bf References}

\ref
Ajhar, E. A.,  Blakeslee, J. P., Tonry, J. L. 1994, \aj 108, 2087

\ref
Ashman, K. M., Zepf, S. E. 1992, \apj 384, 50

\ref
Conti, P. S., Vacca, W. D. 1994, \apj 423, L97

\ref
Couture, J., Harris, W. E., Allwright, J. W. B. 1990, \apjs 73, 671

\ref
Driver, S. P., Windhorst, R. A., Ostrander, E. J., Keel, W. C.,
Griffiths, R. E., R. E., Ratnatunga, K. U., 1995, \apjl submitted and
revised

\ref
Flemming, D., Harris, W., Pritchet, C., Hanes, D., 1995 \aj 109, 1044

\ref
Freedman, W. L., \etal 1994 \nature 371, 757

\ref
Glazebrook, K., Ellis, R., Santiago, B. X., Griffiths, R. E.,
1995, \nature submitted

\ref
Griffiths, R., \etal 1994, \apj 437, 67

\ref
Harris, H., Baum, W., Hunter, D., Kreidl, T., 1991, \aj 101, 677

\ref
Harris, W. E. 1986, \aj 91, 822

\ref
Harris, W. E. 1991, \araa 29, 543

\ref
Harris, W. E. 1995, in IAU Symp. 164

\ref
Harris, W. E.,  Pudritz, R. E. 1994 \apj 429, 177

\ref
Holtzman, J. A., \etal 1992, \aj103, 691

\ref
Holtzman, J. A., \etal 1995, preprint

\ref
Irwin, M. 1985, \mnras 214, 575

\ref
Lee, M. G., Geisler, D. 1993, \aj 106, 493

\ref
Lutz, D. 1991, \aa 245, 31

\ref
Pierce, M., Welch, D., McClure, R., van den Bergh, S., Racine, R.,
Stetson, P.,

1994 \nature 371, 385

\ref
Santiago, B. X., Gilmore, G., Elson, R. A. W. 1995, in preparation

\ref
Whitmore, B., Schweizer, F., Leitherer, C., Borne, K., Robert, C. 1993,

\aj 106, 1354

\ref
Zepf, S. E., Ashman, K. M. 1993, \mnras 264, 611

\vfill\eject
\noindent
{\bf Figure Captions}

\def\fig #1, #2, #3, #4 {\smallskip\leftskip2.5em \parindent=0pt
#4}
%The following macro will give inserted figures. Comment it out if you
%don't have psfig.
\input psfig
\def\fig #1, #2, #3, #4 {
\topinsert
\smallskip
\centerline{\psfig{figure=#1,height=#2 in,width=#3 in}}
\medskip
{\smallskip\leftskip2.5em \parindent=0pt
#4}
\endinsert}

\fig 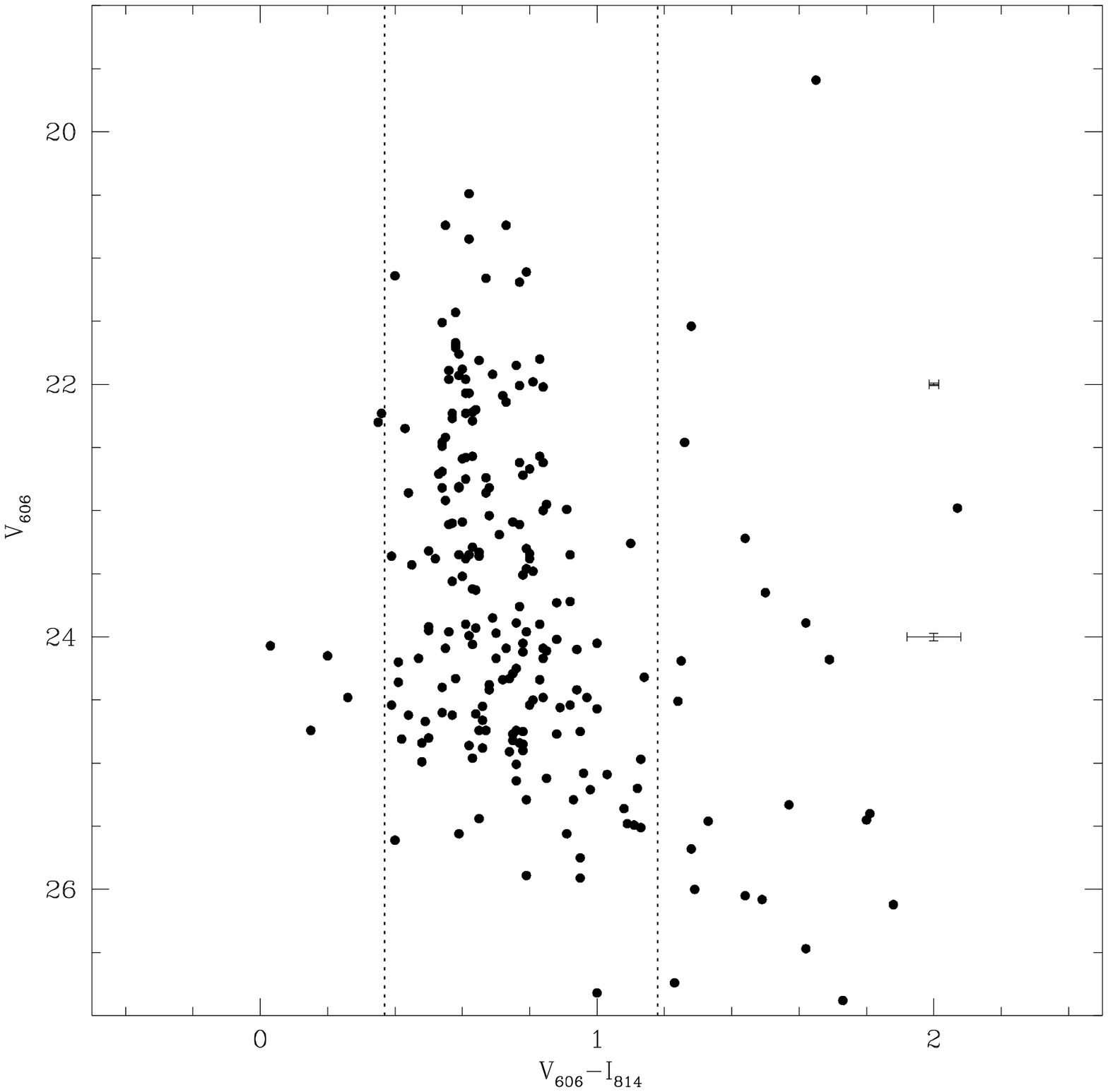, 5, 5, {2- Colour-magnitude diagram with
instrumental HST colours
for 238 objects.  The vertical
dashed lines indicate the limits of $(V_{606}-I_{814})$ for globular
clusters in M87, as discussed in Section 2.ii.  Representative Poisson
error bars are shown.  No reddening corrections have been applied. }

\fig 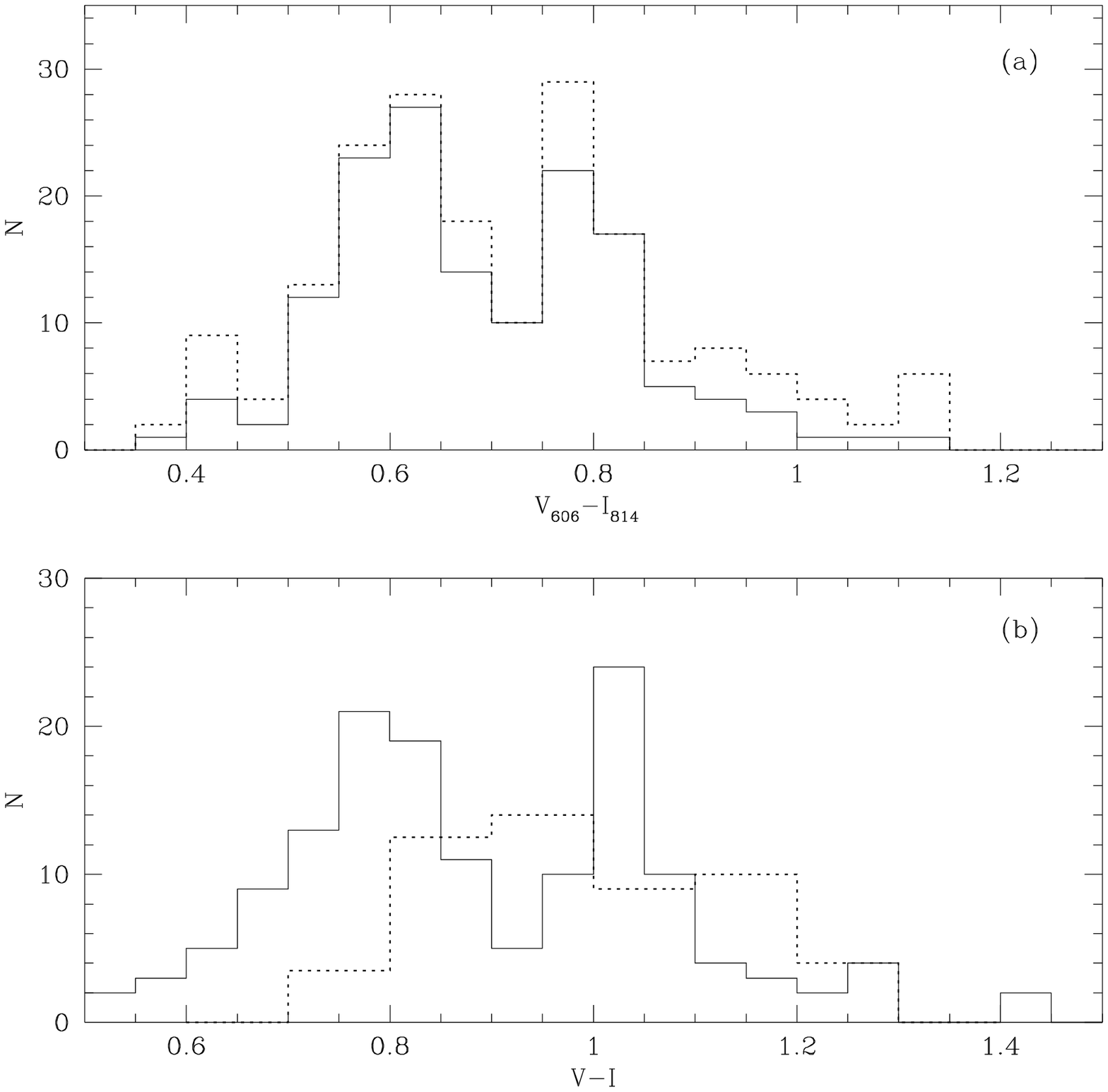, 5, 5, {3- (a) ($V_{606}-I_{814}$) distributions for the 187
globular cluster
candidates selected by colour (dotted line), and for the final sample
of 147 globular clusters, from which
galaxies have been eliminated on the basis of morphology (solid line).  (b)
($V-I$) colour distributions for the 147  globular clusters in our final
sample (solid line), compared with the distribution from Couture \etal for
106 M87 clusters
with $V < 23.5$ (dotted line).
Our distribution appears to be shifted
bluewards by $\approx 0.15$ mag. }

\fig 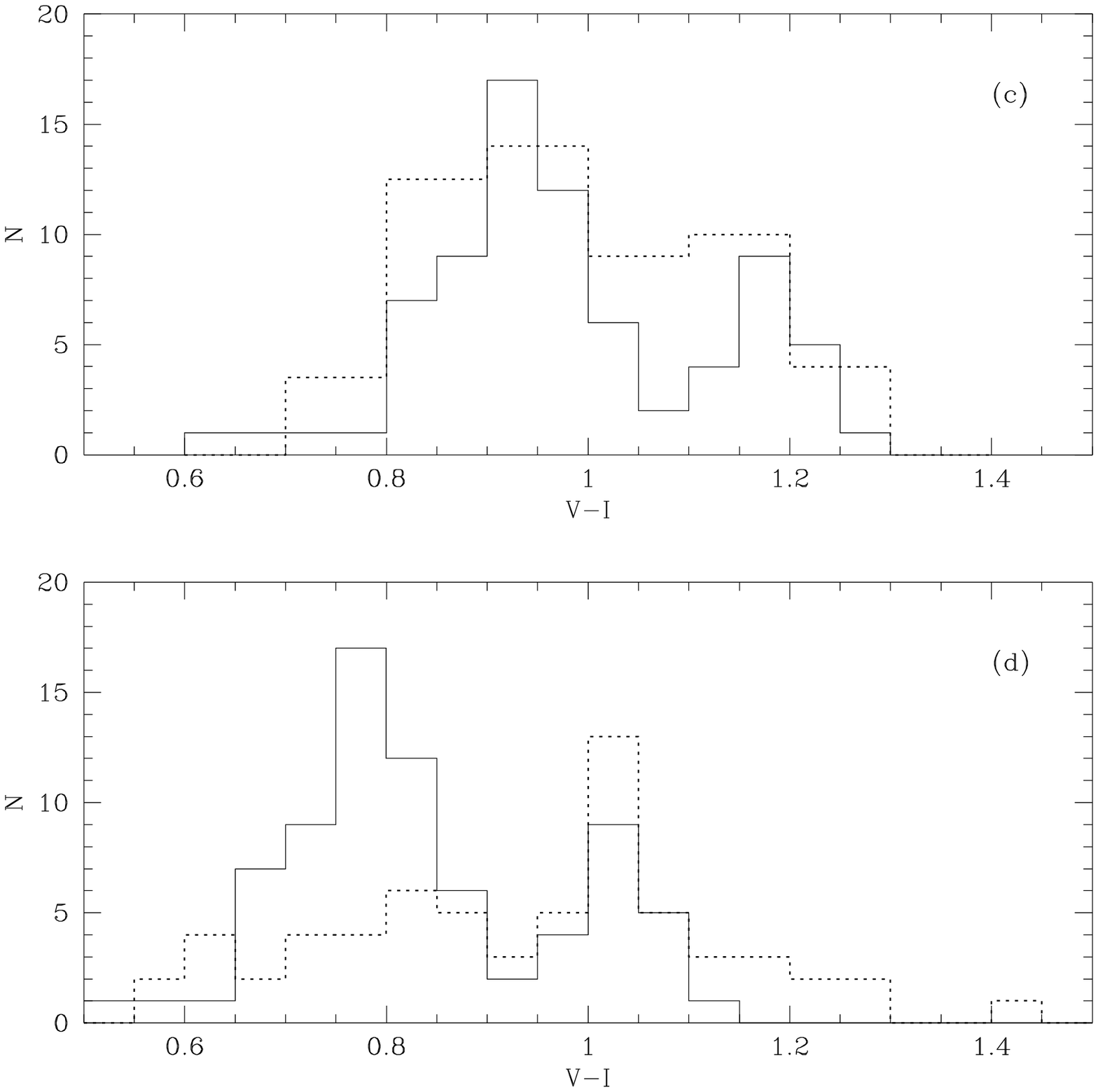, 5, 5, {3 (cont.) - (c) ($V-I$) colour
distributions for the 76  brightest
globular clusters ($V < 23.5$) in our final
sample (solid line), compared with that from Couture \etal (1990).
Our histogram has been reddened by 0.15 mag to bring
it into coincidence with that of Couture \etal
(d) Histograms for the 76 brightest clusters ($V < 23.5$)
(solid line) and
64 fainter clusters ($23.5 < V < 25$) (dotted line) in our final sample.
No colour offsets have been applied. }

\fig 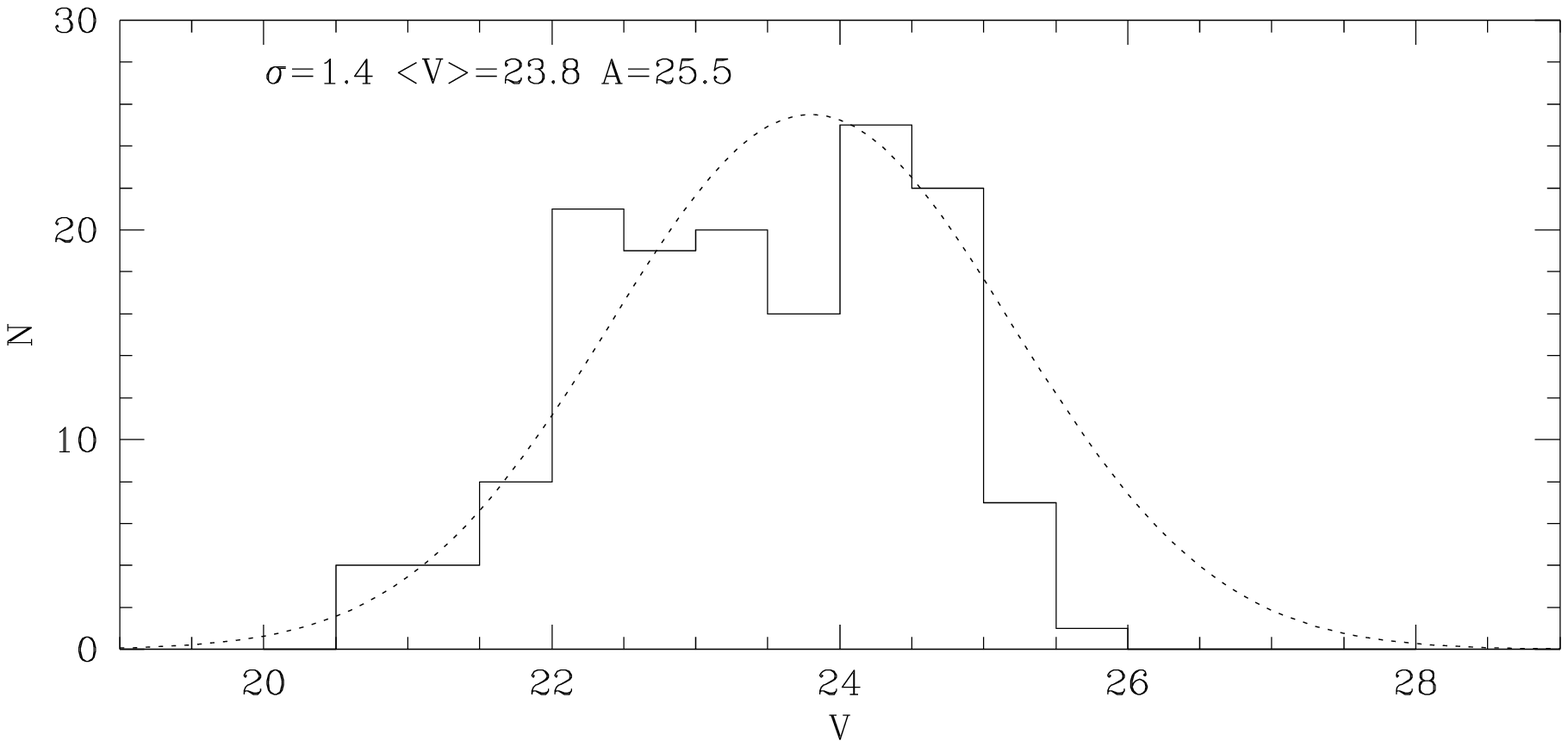, 5, 5, {4- Luminosity function for the 147 clusters in our final
sample.  Superposed is a gaussian distribution with $<V>=23.8$ ($<M_V>=-7.2$)
and $\sigma=1.4$, normalized to the number of clusters with
$V<24$ ($N=100$). }

%{\bf Figure 1.}  Subsection (25.5 x 25.5 arcsec) of the coadded
%F606W image.  The objects in the image range in magnitude from $V\sim21$
%to $V\sim25$.  For example, the bright cluster near the top right
%corner has $V=21.0$, and the close pair of clusters near the bottom centre
%have $V=22.3$ and 23.5.
%Several galaxies in the image are clearly distinguishable from the
%compact globular cluster candidates by their morphology and
surface brightness.
\end